\documentclass[12pt] {article}
\pdfoutput=1

\usepackage{graphicx}
\usepackage{amsmath}
\usepackage{caption}
\usepackage{tabularx}
\usepackage{amssymb}
\usepackage{makecell}
\usepackage{algorithm}
\usepackage[noend]{algpseudocode}
\usepackage{pdfpages}

\usepackage[obeyspaces,hyphens,spaces]{url}
\usepackage[english]{babel}
{\begin{list}{}%
         {\setlength{\leftmargin}{#1}}%
         \item[]%
}
{\end{list}}

\begin{document}
\sloppy

\title{\Large{\textbf{ETMA: A New Software for Event Tree
			Analysis with Application to Power Protection} }}
\author{Mohamed Abdelghany, Waqar Ahmad, Sofi\`ene Tahar, and Sowmith Nethula \vspace*{2em}\\
Department of Electrical and Computer Engineering,\\
Concordia University, Montr\'eal, QC, Canada 
\vspace*{1em}\\
\{m\_eldes,waqar,tahar,s\_nethul\}@ece.concordia.ca
 \vspace*{3em}\\
\textbf{TECHNICAL REPORT}
\vspace*{1em}\\
June 2020
\date{}
}

\maketitle

\newpage
\begin{abstract}
	
	Event Tree (ET) analysis is a widely used forward deductive safety analysis technique for decision-making at a system design stage. Existing ET tools usually provide Graphical Users Interfaces (GUI) for users to manually draw system level ET diagrams, which consist of nodes and branches, describing all possible success and failure scenarios. However, these tools do not include some important ET analysis steps, e.g., the automatic generation and reduction of a complete system ET diagram. In this paper, we present a new \textit{Event Trees Modeling and Analysis} ($\mathcal{ETMA}$) tool to facilitate users to conduct a complete ET analysis of a given system. Some key features of $\mathcal{ETMA}$ include: (i)  automatic construction of a complete ET model of real-world systems; (ii) deletion/reduction of unnecessary ET nodes and branches; (iii)  partitioning of ET paths; and (iv) probabilistic analysis of the occurrence of a certain event. For illustration purposes, we utilize our  $\mathcal{ETMA}$ tool to conduct the ET analysis of a protective fault trip circuit in power grid transmission lines. We also compared the $\mathcal{ETMA}$ results with Isograph, which is well-known commercial tool for ET~analysis.
	
\end{abstract}



\section{Introduction}

Nowadays, the fulfillment of stringent safety requirements for highly critical systems, which are prevalent, e.g., in smart grids and autonomous systems, has been encouraging safety design engineers to utilize dependability analysis techniques as per recommendations of the safety standards, such as IEC 61850 \cite{mackiewicz2006overview} and ISO 26262~\cite{palin2011iso}. Event Tree (ET) analysis is a well-known dependability analysis technique that enumerates all possible combinations of component states and external events and thus provides a detailed system view~\cite{ferdous2009handling}. The building of a graphical diagram of a system ET model starts with an initiating node and sequentially drawing all the system components and their operating states \cite{hu1999evaluating}. In the ET analysis, the probabilistic assessment of the occurrence of a certain event can be used for decision-making at the systems design stage. The results of the ET analysis are extremely useful for safety analysts to quantify systems improvement. \\

Existing commercially available ET tools, such as ITEM~\cite{ITEM_tp}, Isograph \cite{Isograph_tp}, and EC Tree~\cite{ECTREE_tp}, provide many powerful features, including user-friendly editors, a commonly used events library and the coloring of diagram elements for easier viewing. For instance, the EC Tree tool, which is developed by NASA's IM\&S (Integrated Modeling and Simulation) Team, provides an Excel sheet for potential users. It can be easily used to input a given system ET model with a little training. All these tools require a system ET diagram from the user, which is then followed by assigning the probability to each branch of an ET diagram. Prior to utilizing these tools for ET analysis, the users must draw a given system ET diagram manually, may be on a paper. However, this manual approach may introduce errors from the start since an ET diagram becomes significantly large as the number of system components and their operational states increases. Moreover, an important feature of partitioning an ET with respect to an event occurrence and then to calculate its corresponding probability is not available in any existing ET analysis tools. \\

To overcome the above-mentioned limitations of existing ET tools, we develop a new Event Trees Modeling and Analysis ($\mathcal{ETMA}$) tool.  It is mainly inspired from the work of Papazoglou \cite{papazoglou1998mathematical}, who was among the first ones to describe the sound mathematical foundations of ET analysis during late 90's. The development of $\mathcal{ETMA}$ starts from a recursive function describing the pattern of generating an ET diagram from the given list of all possible failure and success modes of given system components. Most importantly, $\mathcal{ETMA}$ offers a reduction feature, which deletes unnecessary nodes and branches from the automatically generated ET diagram and return an ET model representing the actual behavior of a given system. $\mathcal{ETMA}$ has  an  intriguing  feature  of  partitioning the ET paths according to the system components failure and success modes. It also provides the probabilistic analysis feature by allowing users to assign the probability to each components states. Moreover,  the $\mathcal{ETMA}$ results can be used to identify critical components and make decisions about adding redundancy in a system. All these $\mathcal{ETMA}$ features are developed in the Python programming language~\cite{van2011python}, which offers extensive built-in libraries for displaying, list manipulations and arithmetic~calculations. \\

It is worth mentioning that our $\mathcal{ETMA}$ tool can handle large and complex real-world systems with an arbitrary number of system components and their operating states. For illustration purposes, we utilize $\mathcal{ETMA}$ to conduct the ET analysis of a protective trip circuit in power grid transmission lines consisting of several critical components, such as relays and current transformers\cite{grainger2003power}.\\

Our main novel contributions in this paper are as follows:

\begin{itemize}
	
	\item Automatic generation of complete system ET model from a given list of system components and their operating states 
	\item Deletion of unnecessary nodes and branches to generate a reduced ET model 
	\item Partitioning of ET with respect to an event occurrence for probabilistic analysis
	\item Implementation in Python of a comprehensive tool for ET modeling and analysis: $\mathcal{ETMA}$
	\item Step-wise ET analysis of a protection fault trip circuit in power grid transmission lines with a decision analysis to add redundancy for some critical components
	\item Comparison between the results of $\mathcal{ETMA}$ with the commercial Isograph ET analysis software
	
\end{itemize}

The rest of the paper is organized as follows: In Section~II, we briefly summarize the fundamentals of ETs and the theoretical foundations of the $\mathcal{ETMA}$ tool. Section III describes the ET modeling and analysis features of $\mathcal{ETMA}$. Section~IV presents the step-wise process of ET analysis of a protective trip circuit in power grid transmission lines using $\mathcal{ETMA}$ and the decision analysis of the trip circuit critical components based on $\mathcal{ETMA}$ results. Section V provides a comparison between $\mathcal{ETMA}$ and the Isograph~software. Lastly, Section VI concludes the paper.


\section{Event Trees}

An ET diagram starts by a single initiating event called \textit{Node} and then all possible outcomes of an event are drawn as branches.  This process is continuously repeated in the forward direction until all event nodes and their branches are drawn resulting in a complete ET diagram of the system. Fig. 1 depicts a generic ET diagram.\\

\begin{figure}
	\centering 
	\includegraphics[width = 0.83 \columnwidth]{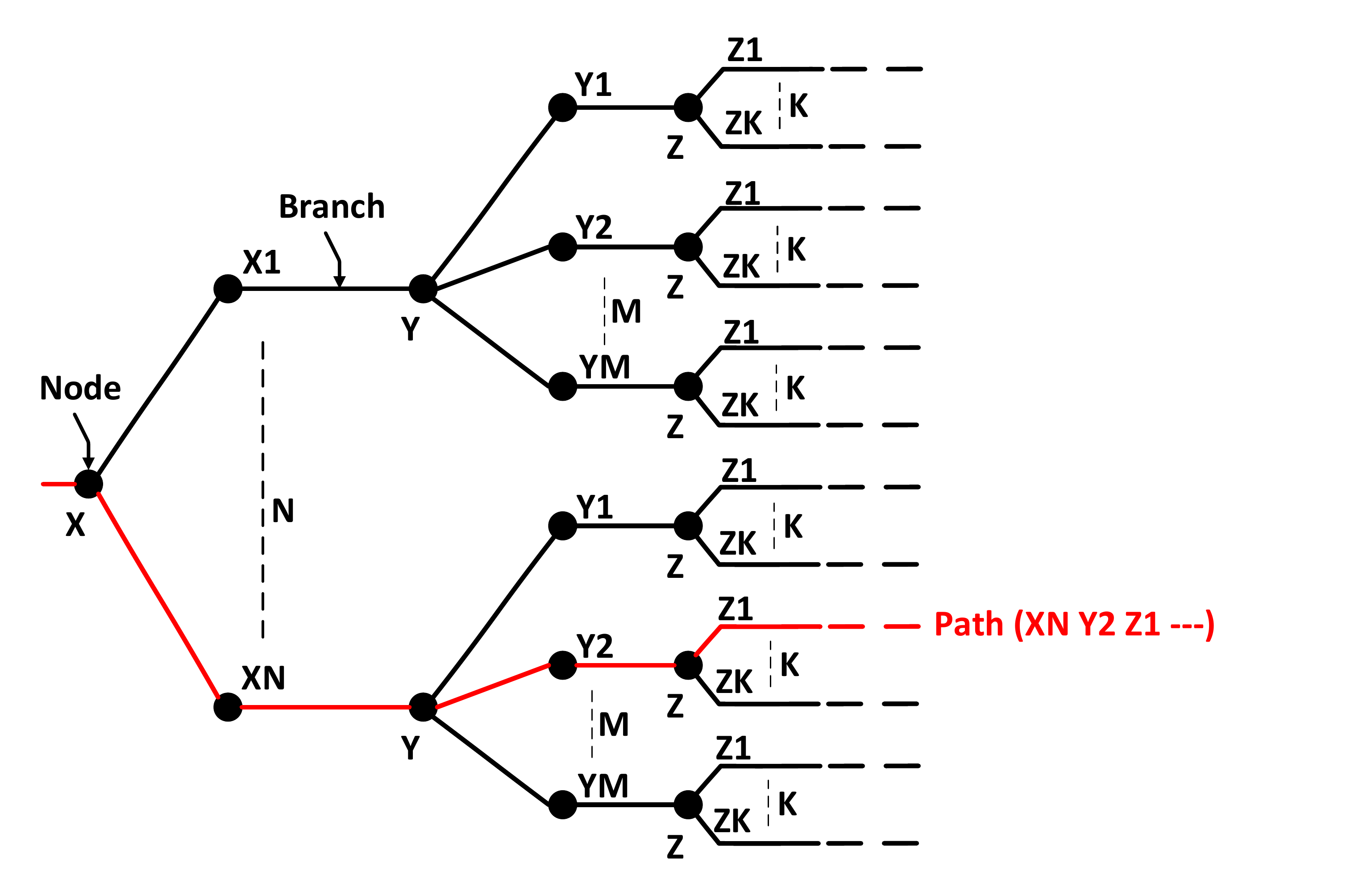} 
	\caption{\protect A generic ET diagram}
	\label{Fig: Event Tree}
\end{figure}

Nodes model the occurrence of different possibilities for an event or modes of operation of system components, which is known as event outcome space in the literature~\cite{papazoglou1998mathematical}. A \textit{Node} is usually represented by a circle with multiple line segments. For instance, in Fig. \ref{Fig: Event Tree}, X, Y and Z are nodes. Branches originating from a node represent each of the next possible component states. A \textit{Branch} is usually designated by a line segment associated with a preceding node. For instance, {$X_1$,\dots, $X_N$} and {$Y_1$,\dots, $Y_M$} are branches, as shown in Fig. \ref{Fig: Event Tree}. A complete ET diagram draws all possible paths that represent a specific system. Each path consists of a unique sequence of events, i.e., ($X_N Y_2 Z_1$~\dots) is one of the ET paths in Fig. \ref{Fig: Event Tree}. \\

The probability of each path in an ET diagram is evaluated for decision-making at the systems design stage. These probabilities represent the likelihood of each outcome or condition that can happen in a system. The assessment of these probabilities depends upon the occurrence of previous events in an ET. The probability of each path is usually computed by multiplying the probabilities of events associated with all nodes in a path. For example, the probability of the path ($X_N Y_2 Z_1$~\dots) in Fig. \ref{Fig: Event Tree} is expressed mathematically as: 
\begin{equation}
\centering 
\mathcal{P} (X_N Y_2 Z_1~\dots) = \mathcal{P} (X_N) * \mathcal{P} (Y_2) * \mathcal{P} (Z_1) * \dots
\end{equation}
\noindent Also, all events in a path including the initiating node are assumed to be mutually exclusive. This implies that the cumulative probability of all branches connected to a certain node must be equal to 1 as: 
\begin{equation}
\centering 
\sum_{i=1}^{N} \mathcal{P}(X_i) = 1, \sum_{j=1}^{M} \mathcal{P}(Y_j) = 1, \sum_{h=1}^{K} \mathcal{P}(Z_h) = 1, \dots
\end{equation}

\subsection{Theoretical Foundation}

The underline mathematics of ET analysis in $\mathcal{ETMA}$ are mainly inspired from the work of Papazoglou \cite{papazoglou1998mathematical} that are briefly described as follows: \\

An event outcome space ($\mathcal{W}$) is referred to as a list of all possible outcomes of an event. Each node of an ET is associated with an event outcome space must satisfy following constraints

\begin{enumerate}
	\item \textit{Distinct}: All outcomes in an event outcome space must be unique.
	\item \textit{Disjoint (mutually exclusive)}: Any pair of events from a set of events outcome space cannot occur at the same~time.
	\item \textit{Complete}: An event outcome space must contain all possible events that can occur. 
	\item \textit{Finite}: An event outcome space must consists of a finite number of elements.
	\begin{equation}
	\centering 
	\mathcal{W}= [\omega_j] \qquad j = 1,2,\dots,\mathcal{N}
	\end{equation}
\end{enumerate}

Consider a system having two events, say $\textit{E}_1$ and $\textit{E}_2$, with two event outcome spaces $\mathcal{W}_1$ and $\mathcal{W}_2$, respectively. The Cartesian product ($\bigotimes$) of these event outcome spaces returns a list of ($\mathcal{N}_1 \times \mathcal{N}_2$) pairs containing all possible outcome pairs for the occurrence of $\textit{E}_1$ and $\textit{E}_2$ together (i.e., $\mathcal{W}_1 \bigotimes \mathcal{W}_2$). In ET, the resulting event outcome space from the Cartesian product of two event outcome spaces must also satisfy the above-mentioned constraints. We program this concept in $\mathcal{ETMA}$ in two steps as~follows:

\noindent\textit{Step} 1: We first construct a list of pairs by taking each element from the event outcome spaces $\mathcal{W}_1$ and $\mathcal{W}_2$.\\

\noindent\textit{Step} 2: We ensure that the obtained duets from Step 1 are mutually exclusive. For instance, consider two arbitrary outcomes ($\omega_{1m}$ $\omega_{2n}$) and ($\omega_{1k}$ $\omega_{2l}$), at least ($\textit{m}$ $\neq$ $\textit{k}$) or ($\textit{n}$~$\neq$~$\textit{l}$) must be true.\\

One of our main objectives, in this work, is to take an arbitrary list of given system components with their operating states and automatically generate the corresponding ET diagram (i.e., $\mathcal{W}_1 \bigotimes \mathcal{W}_2 \bigotimes \dots \bigotimes \mathcal{W}_\mathcal{N}$). For this purpose, we developed a Python function in that can recursively perform \textit{Steps} 1 and 2 on a given list of event outcome spaces representing the system components and their operational~states.

To present a clear understanding of the above-mentioned automatic ET generation feature of $\mathcal{ETMA}$,  consider a system having three events, say $\textit{E}_1$, $\textit{E}_2$ and $\textit{E}_3$, with three event outcome spaces $\mathcal{W}_1$, $\mathcal{W}_2$ and $\mathcal{W}_3$, respectively. The resulting ET diagram is shown in Fig. \ref{Fig: Event Trees EXAMPLE} and the collection of all possible ET paths in a list of strings are as: 

\begin{flushleft}
	\hspace{2mm} $Path_0$ \hspace{1.5mm} = [$\textit{A}_1$,  $\textit{B}_1$, $\textit{C}_1$],   
	$Path_1$ \hspace{1mm} = [$\textit{A}_1$, $\textit{B}_1$, $\textit{C}_2$], \\
	\hspace{2mm} $Path_2$ \hspace{1.5mm} = [$\textit{A}_1$, $\textit{B}_2$, $\textit{C}_1$],
	$Path_3$ \hspace{1mm} = [$\textit{A}_1$, $\textit{B}_2$, $\textit{C}_2$], \\
	\hspace{2mm} $Path_4$ \hspace{1.5mm} = [$\textit{A}_2$, $\textit{B}_1$,  $\textit{C}_1$],
	$Path_5$ \hspace{1mm} = [$\textit{A}_2$, $\textit{B}_1$, $\textit{C}_2$], \\
	\hspace{2mm} $Path_6$ \hspace{1.5mm} = [$\textit{A}_2$, $\textit{B}_2$, $\textit{C}_1$],
	$Path_7$ \hspace{1mm} = [$\textit{A}_2$, $\textit{B}_2$, $\textit{C}_2$], \\
	\hspace{2mm} $Path_8$ \hspace{1.5mm} = [$\textit{A}_3$, $\textit{B}_1$, $\textit{C}_1$], 
	$Path_9$ \hspace{1mm} = [$\textit{A}_3$, $\textit{B}_1$, $\textit{C}_2$], \\
	\hspace{2mm} $Path_{10}$ = [$\textit{A}_3$, $\textit{B}_2$, $\textit{C}_1$], 
	$Path_{11}$ = [$\textit{A}_3$, $\textit{B}_2$, $\textit{C}_2$]
\end{flushleft}

\noindent The order of the outcomes in a path is irrelevant when evaluating the probabilities of a given path \cite{papazoglou1998functional}, i.e., the probability of path [$\textit{A}_3$, $\textit{B}_1$,  $\textit{C}_2$] is  $\mathcal{P}$ ($\textit{A}_3$) * $\mathcal{P}$ ($\textit{B}_1$) * $\mathcal{P}$ ($\textit{C}_2$), which is exactly equivalent to the probability of path [$\textit{C}_2$,~$\textit{B}_1$,~$\textit{A}_3$] due to the commutative property of multiplication. However, in many cases, it is useful to preserve the order of outcomes in the ET paths. For instance, it can facilitate the thinking process in certain critical situations, but it has no relation to the dynamic of the system components \cite{papazoglou1998mathematical}. Another benefit of introducing a sequence is that, in some critical systems, if the main component fails, then the probability of this path depends on the failure of the main component only without considering the next components state. Therefore, we believe that a sequence preserving generation of ETs must be~adopted. 


\subsection{Branch or Node Deletion}

During ET analysis, we may require to model the exact logical behavior of systems in the sense that the irrelevant nodes and branches should be removed from a complete ET of a system. This can be done by deleting some specific branch or nodes corresponding to the occurrence of certain events, which are known as \textit{Complete Cylinders}~(CCs). These cylinders are ET paths consisting of $\mathcal{N}$  events and they are conditional on the occurrence of $\mathcal{K}$  events in their respective paths and not conditional on the occurrence of the remaining ($\mathcal{N}$~-~$\mathcal{K}$) events \cite{papazoglou1998functional}. These cylinders are also referred to as CCs  with respect to $\mathcal{K}$  \textit{Conditional Events} (CEs). 

A reduced ET can be obtained in two ways: (1)~eliminate certain branches with all their successor nodes; and (2)~delete only nodes from specific branches leaving the successor nodes connected to these branches. The reduction process can be explained as follows:

\subsubsection{Branch Deletion}
If the paths \{8; 9; 10;~11\}, shown in Fig.~\ref{Fig: Event Trees EXAMPLE}, are CCs with respect to the event $\textit{A}_3$ (i.e., not conditional on the occurrence of neither $\mathcal{W}_2$ nor $\mathcal{W}_3$ event outcome spaces), then the branches [$\textit{A}_3$, $\textit{B}_1$] and [$\textit{A}_3$, $\textit{B}_2$] should be deleted. The resulting ET paths after deletion are as~follows:

\begin{flushleft}
	\hspace{2mm} $Path_0$ = [$\textit{A}_1$,  $\textit{B}_1$, $\textit{C}_1$],   
	$Path_1$ = [$\textit{A}_1$, $\textit{B}_1$, $\textit{C}_2$], \\
	\hspace{2mm} $Path_2$ = [$\textit{A}_1$, $\textit{B}_2$, $\textit{C}_1$],
	$Path_3$ = [$\textit{A}_1$, $\textit{B}_2$, $\textit{C}_2$], \\
	\hspace{2mm} $Path_4$ = [$\textit{A}_2$,  $\textit{B}_1$, $\textit{C}_1$],   
	$Path_5$ = [$\textit{A}_2$, $\textit{B}_1$, $\textit{C}_2$], \\
	\hspace{2mm} $Path_6$ = [$\textit{A}_2$, $\textit{B}_2$, $\textit{C}_1$],
	$Path_7$ = [$\textit{A}_2$, $\textit{B}_2$, $\textit{C}_2$], \\
	\hspace{2mm} $Path_8$ = [$\textit{A}_3$]
	
\end{flushleft}

\subsubsection{Node Deletion}
\begin{figure}[!t]
	\centering
	\includegraphics[width = 0.8 \columnwidth]{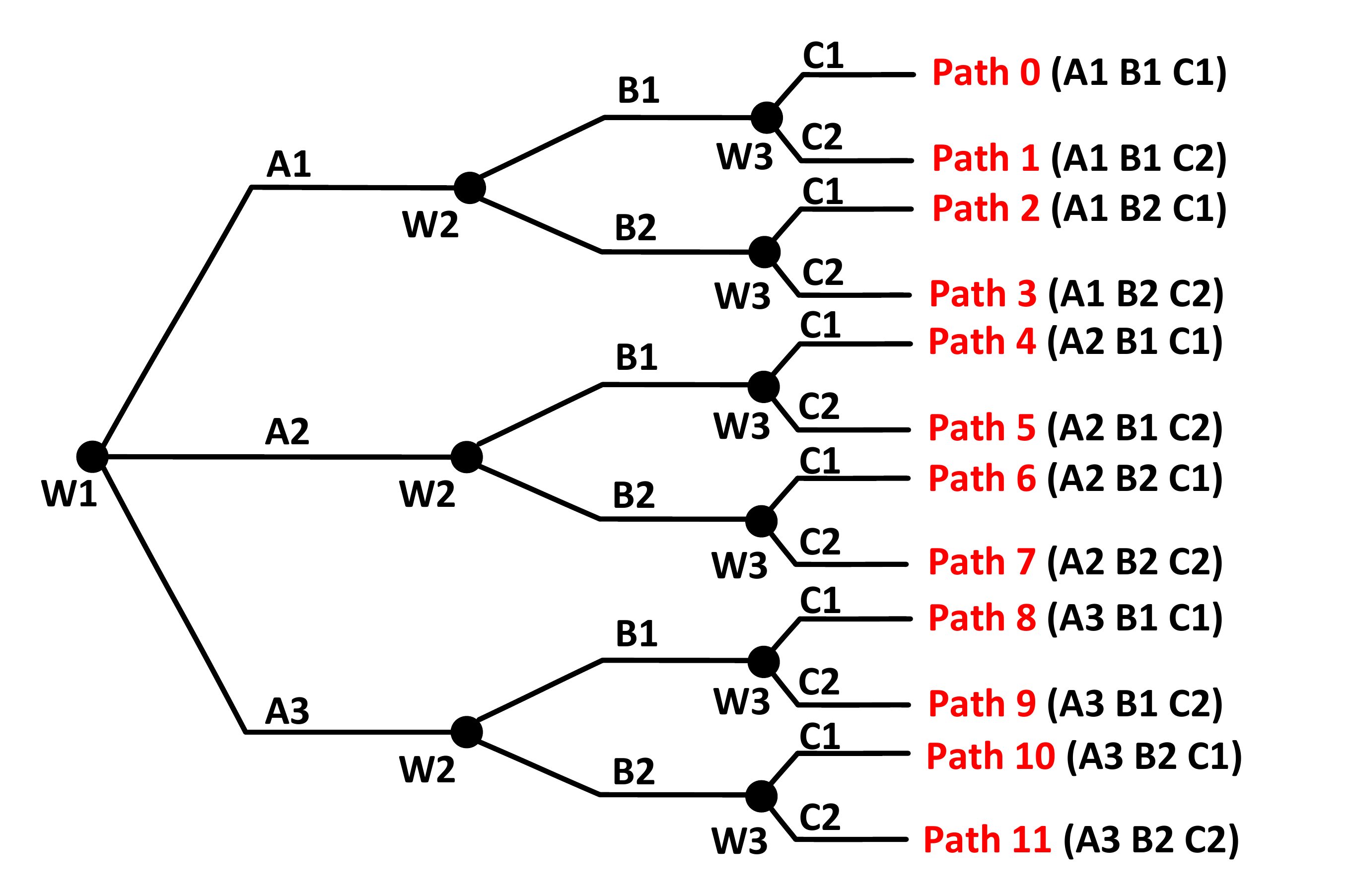} 
	\caption{\protect ET represents the event space outcomes}
	\label{Fig: Event Trees EXAMPLE}
\end{figure}

If the paths \{0; 1; 2;~3\}, shown in Fig.~\ref{Fig: Event Trees EXAMPLE}, are CCs with respect to the event $\textit{A}_1$ and $\mathcal{W}_3$ event outcome space (i.e., not conditional on the occurrence of $\mathcal{W}_2$ event outcome space only), then the node $\mathcal{W}_2$ from the branch $\textit{A}_1$ needs to be deleted only while keeping the $\mathcal{W}_3$ event outcome space connected with the event $\textit{A}_1$. The resulting ET paths after deletion are as~follows:

\begin{flushleft}
	\hspace{2mm} $Path_0$ = [$\textit{A}_1$, $\textit{C}_1$], $Path_1$ = [$\textit{A}_1$, $\textit{C}_2$],\\
	\hspace{2mm} $Path_2$ = [$\textit{A}_2$, $\textit{B}_1$,  $\textit{C}_1$],
	$Path_3$ = [$\textit{A}_2$, $\textit{B}_1$, $\textit{C}_2$], \\
	\hspace{2mm} $Path_4$ = [$\textit{A}_2$, $\textit{B}_2$, $\textit{C}_1$],
	$Path_5$ = [$\textit{A}_2$, $\textit{B}_2$, $\textit{C}_2$], \\
	\hspace{2mm} $Path_6$ = [$\textit{A}_3$]
\end{flushleft}

\section{Event Tree Analysis in $\mathcal{ETMA}$}

\begin{figure}
	\centering
	\includegraphics[width = 1 \columnwidth]{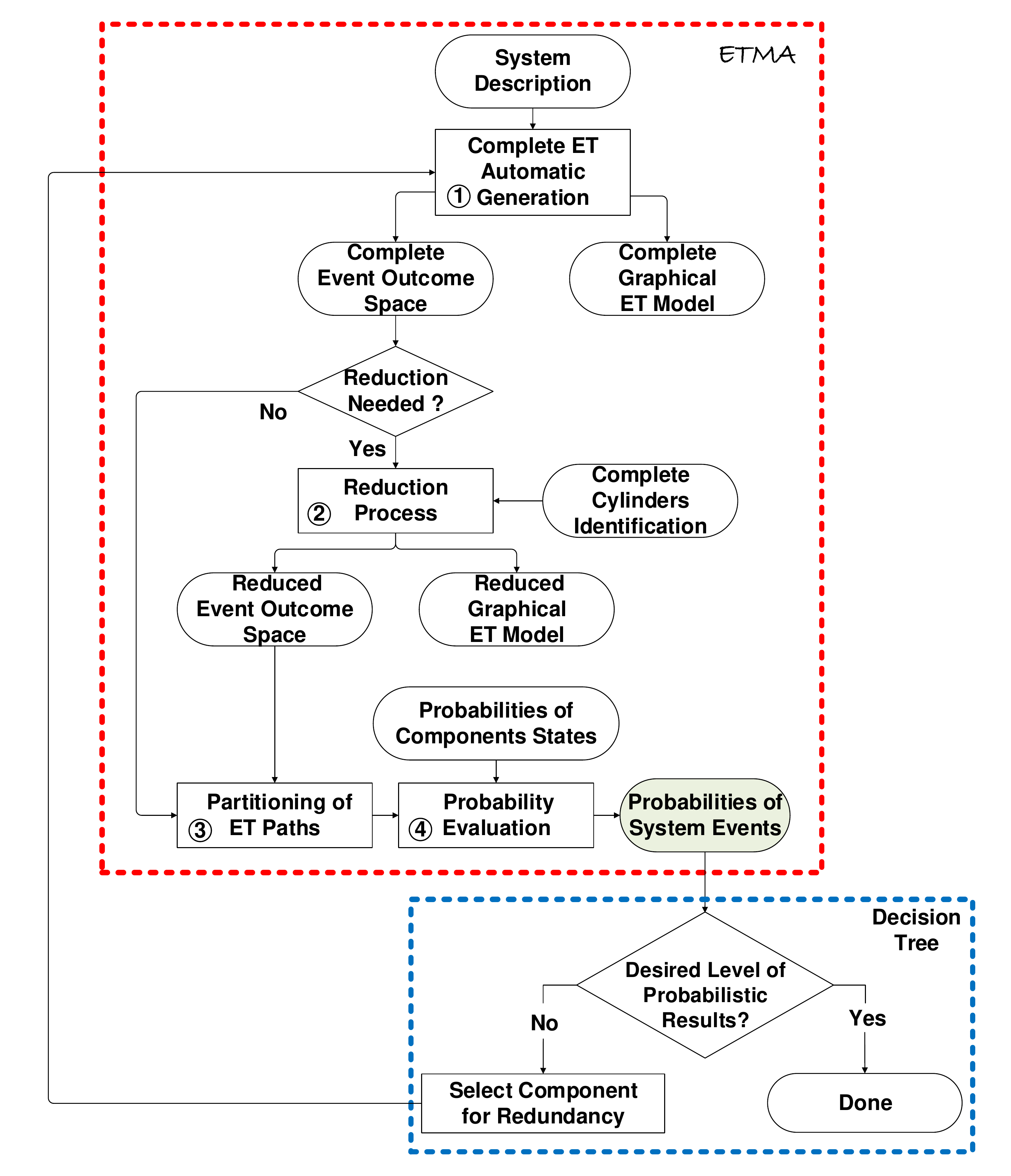} 
	\caption{\protect $\mathcal{ETMA}$ Analysis flowchart}
	\label{Fig: Event Trees Flow chart}
\end{figure}

The flowchart describing the $\mathcal{ETMA}$ tool for ETs modeling and analysis is depicted in Fig.~\ref{Fig: Event Trees Flow chart} and mainly consists of 4 steps as follows: (1) identify the given system components and their operating states representing the behavior of the system, then automatically generate a complete ET model describing all system components states and also produce a complete outcome space with all possible scenarios of different levels of failure and success; (2) optionally, reduce manually some nodes/branches from the generated complete ET diagram to regenerate a smaller model exhibiting the exact behavior of the given system; (3) partition the ET paths according to the system components failure and success modes; and (4) evaluate the probability of occurrence for certain events in the system after partitioning the ET~paths.\\

The details of the $\mathcal{ETMA}$ functions that perform the above-mentioned operations are described in Algorithm \ref{algorithm}. We provide pop-up input windows for each of these functions in order to facilitate the users interaction with the $\mathcal{ETMA}$ tool. It can be noticed from Algorithm \ref{algorithm} that the reduction $\mathcal{ETMA}$ feature can be bypassed, in case the deletion of nodes or branches is not required. Also, to ensure that $\mathcal{ETMA}$ is capable of generating complex and scalable ETs, we have implemented the steps of Algorithm \ref{algorithm} using the \textit{PyGraphviz} Python package~\cite{PyGraphviz_tp}, which provides several methods for layout and drawing of complex graphs.\\

\begin{algorithm}
	\caption{\protect  \centering $\mathcal{ETMA}$}
	\label{algorithm}
	\begin{algorithmic}[1]
		\Procedure{}{}\\
		
		\textbf{S1: complete\_gen} \\
		\hspace{5mm} \textbf{Input:}
		system name \\
		\hspace{16mm} system components\\
		\hspace{16mm} each system component states \\
		\hspace{5mm} \textbf{Output:}
		complete ET model \\
		\hspace{18.5mm} complete event outcome space\\
		\textbf{If}
		Reduction of ET model needed? \\
		\hspace{8mm} \textbf{then}\\
		\hspace{8mm} \textbf{S2: reduction\_process} \\
		\hspace{14mm} \textbf{Input:} CCs identification\\
		\hspace{14mm} \textbf{Output:} reduced ET model\\
		\hspace{27.5mm} reduced event outcome space\\
		\textbf{S3: partitioning\_paths} \\
		\hspace{5mm} \textbf{Input:} component event name(s) \\
		\hspace{16mm} ET path number(s) \\ 
		\hspace{5mm} \textbf{Output:} system events ET paths\\
		\textbf{S4: probability\_eval} \\
		\hspace{5mm}  \textbf{Input:} probabilities of components states \\
		\hspace{5mm} \textbf{Output:} Occurrence probability of an event \\
		\textbf{end procedure}
		
		\EndProcedure
	\end{algorithmic}
\end{algorithm}

In reliability engineering, the decisions to add redundancy for critical components or functions in a system are very crucial since it significantly increases the cost of the system. Redundancy is often used in the form of a backup or fail-safe in order to improve actual system performance. Decision tree~\cite{niuniu2010review}, is a tree-like model that helps safety engineers to conduct decision analysis and make effective decisions, like adding redundancy to critical system components. Fig.~\ref{Fig: Event Trees Flow chart} shows the procedure of making a decision for redundancy of a critical component in a system. If the level of the probabilistic analysis evaluated from the ET model is satisfied, then this component is duplicated. If the results are not acceptable, then another critical component is selected for redundancy from the system and $\mathcal{ETMA}$ is used for re-construct the new ET model. \\

In the next section, we apply our algorithm and tool, which can be downloaded from \cite{ETMA_TP}, on a real-world system for the domain of power protection and the results of $\mathcal{ETMA}$, in detail, are uploaded on the same above link.


\section{Trip Circuit Analysis}
\label{case study}
Consider a trip circuit in a power grid system, which is used to isolate a specific transmission line from the rest of a power grid, in case a fault occurs. The cascaded failure for many transmission lines could lead to a blackout situation for the whole grid, like what happened in San Diego in 2011~\cite{portante2014simulation}. Therefore, a detailed ET analysis of the trip circuit is essential. \\

\begin{figure}
	\begin{center}
		\centering
		\includegraphics[width = 0.87 \columnwidth]{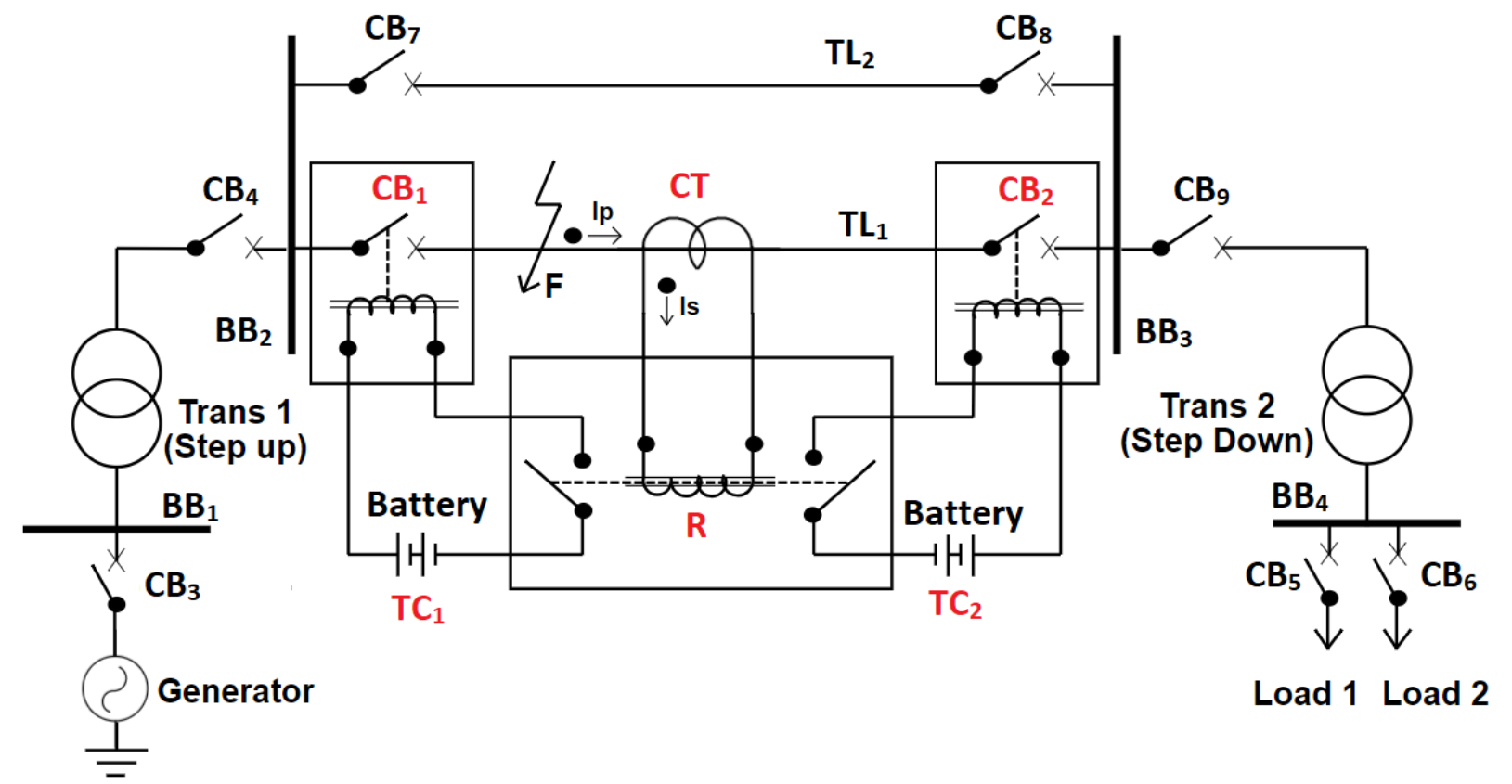} 
	\end{center}
	\caption{\protect Single line diagram of a trip circuit in a power grid}
	\label{Fig: Case study}
\end{figure}

The power grid consists of one generator, 9 circuit breakers~(CB), 4 bus bars (BB), 2 transmission lines (TL), 2 loads, 2 (on step up and one step down) transformers (Trans), 2 trip circuits (TC) with 1 relay (R) and 1 current transformer (CT), as shown in Fig. \ref{Fig: Case study}. During normal operation, all CBs are in a closed position. If a fault (F) occurs on $TL_1$, a primary current $(I_p)$ spike rises to about 20 times from a normal current level. Then, the CT detects that there is a fault in $TL_1$ and the secondary current $(I_s)$ also rises with the same ratio simultaneously. Consequently, the relay coil increases the magnetic field and attracts the relay contacts, which are connected to the two separated trip circuits 1 and~2.  Each trip circuit is provided with a battery. So, when the relay contact closes, it becomes a closed loop. Finally, the magnetic field produced by the trip coils 1 and 2 will push $CB_1$ and $CB_2$ to open and isolate $TL_1$. If all components of the trip circuit work correctly, then the fault becomes isolated and the grid is safe. If not, then the grid is in a risk situation of a blackout and back-up decisions should be made. In this paper, we study the ET-based probabilistic analysis of all scenarios of failure and success that can occur in the trip circuit.

\subsection{Event Tree Analysis}

We start the ET analysis of the trip circuit in $\mathcal{ETMA}$ by first generating a complete ET model. Then, we delete the unnecessary nodes and branches to obtain a reduced ET that models the actual behavior of the trip circuit. Afterwards, we estimate the probabilities of different events that can occur in the trip circuit, for instance, the probability of both breakers $CB_1$ and $CB_2$ failing. Following are the steps required to conduct the trip circuit ET analysis in $\mathcal{ETMA}$:\\

\begin{figure} [!t]
	\begin{center}
		\centering
		\includegraphics[width = 0.74 \columnwidth]{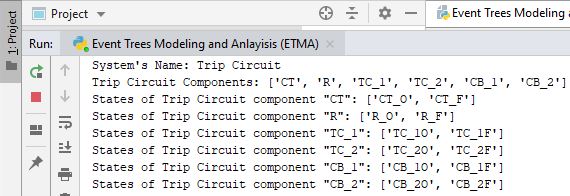} 
	\end{center}
	\caption{\protect $\mathcal{ETMA}$: Trip circuit identification  O (Operates) / F (Fails to operate)}
	\label{Trip Circuit identification}
\end{figure}

\begin{figure} [!t]
	\begin{center}
		\centering
		\includegraphics[width = 0.74 \columnwidth]{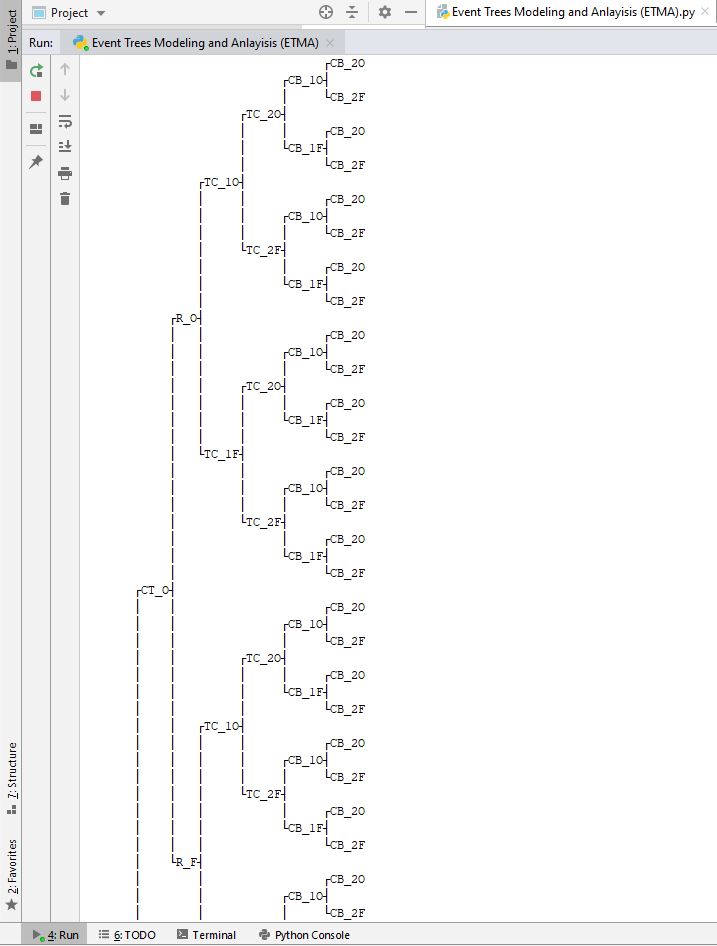} 
	\end{center}
	\caption{\protect $\mathcal{ETMA}$: Trip circuit complete ET model}
	\label{Fig: ETAT complete ET generation}
\end{figure}

\noindent\textit{Step 1 (Complete ET Generation)}: \\
We enter the details of the trip circuit components consisting of one~$CT$, one $R$, two TCs ($TC_1$ and $TC_2$) and two CBs ($CB_1$ and $CB_2$) and each having two operational states, i.e., operating or failing, as shown in Fig.~\ref{Trip Circuit identification}. However, we can assign different levels of failure associated with each component. The entered details are sufficient for $\mathcal{ETMA}$'s function to automatically generate the complete graph ET model, see Fig. \ref{Fig: ETAT complete ET generation} for a snapshot of a portion of the complete ET. This model shows the whole  possible scenarios of failure and success for the trip circuit components states. $\mathcal{ETMA}$ also automatically produces a complete event outcome space (64 paths from 0 to 63) from the complete ET model as: 
\begin{flushleft}
	\hspace{2mm} $Path_0$ = [$CT_O$, $R_O$, $TC_{1O}$, $TC_{2O}$, $CB_{1O}$, $CB_{2O}$] \\
	\hspace{2mm} $Path_1$ = [$CT_O$, $R_O$, $TC_{1O}$, $TC_{2O}$, $CB_{1O}$, $CB_{2F}$]\\
	\hspace{5cm} \vdots \\
	\hspace{2mm} $Path_{62}$ = [$CT_F$, $R_F$, $TC_{1F}$, $TC_{2F}$, $CB_{1F}$, $CB_{2O}$]\\
	\hspace{2mm} $Path_{63}$ = [$CT_F$, $R_F$, $TC_{1F}$, $TC_{2F}$, $CB_{1F}$, $CB_{2F}$]
\end{flushleft}

\begin{figure}[!t]
	\begin{center}
		\centering
		\includegraphics[width = 0.74 \columnwidth]{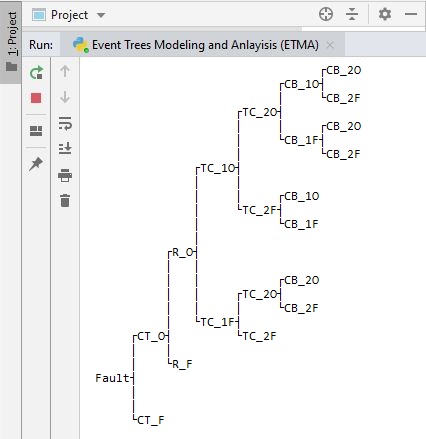} 
	\end{center}
	\caption{\protect $\mathcal{ETMA}$: Trip circuit reduced ET model}
	\label{Fig: ET case study}
\end{figure}

\noindent\textit{Step 2 (ET Reduction Process)}: \\
If the user desires to take into consideration the complete ET model generated in \textit{Step}~1, then $\mathcal{ETMA}$ provides a bypassing option for \textit{Step} 2 (i.e, ET reduction~process). Otherwise, the next step is to define the CCs and their CEs ~(Table~\ref{Table: ET Complete Cylinders of Trip circuit}) to model the exact logical behavior of the trip circuit system. For instance, consider the paths from 32 to 63, if the CT fails then the likelihood or probability of occurrence of these paths are equal to the probability of CT failure only, regardless of the status of other components (i.e, the paths from 32 to 63 are CCs with respect to $CT_F$). So, in $\mathcal{ETMA}$, we deleted the branches [$CT_F$, $R_O$] and [$CT_F$,~$R_F$] from the complete ET (64 paths) in order to model the exact logical behavior of the trip circuit, as shown in Fig.~\ref{Fig: ET case study}. The reduced event outcome space (11 paths from 0 to~10) produced from the reduced ET model is as: \\

\begin{table} [!t]
	\caption{\protect  \centering Trip circuit ET complete cylinders}
	\label{Table: ET Complete Cylinders of Trip circuit}
	\begin{center}
		\begin{tabular}{|c|c|l|c|} 
			\hline \thead{CCs} & \thead{ET Paths} & \thead{CEs} & \thead{\makecell{Type of \\Deletion}} \\ [0.5ex] 
			\hline\hline
			\makecell{$CC_1$}  & \makecell{32,\dots,63}  & $CT_F$ & \makecell{Branch} \\  [1ex]
			\hline
			\makecell{$CC_2$}  & \makecell{16,\dots,31} & $CT_O$, $R_F$ & \makecell{Branch } \\ [1ex]
			\hline 
			\makecell{$CC_3$}  & \makecell{12,\dots,15} & $CT_O$, $R_O$, $TC_{1F}$, $TC_{2F}$ & \makecell{Branch } \\ [1ex]
			\hline
			\makecell{$CC_4$}  & \makecell{8,\dots,11}  & $CT_O$, $R_O$, $TC_{1F}$, $TC_{2O}$, $CB_{2}$ & \makecell{Node ($CB_{1}$)} \\ [0.5ex]
			\hline
			\makecell{$CC_5$}  & \makecell{4,\dots,7} &  $CT_O$, $R_O$, $TC_{1O}$, $TC_{2F}$, $CB_{1}$ & \makecell{Branch } \\ [0.5ex]
			\hline
		\end{tabular}
	\end{center}
\end{table} 

\begin{flushleft}
	\hspace{2mm} $Path_0$  = [$CT_O$, $R_O$, $TC_{1O}$, $TC_{2O}$, $CB_{1O}$, $CB_{2O}$]\\
	\hspace{2mm} $Path_1$  = [$CT_O$, $R_O$, $TC_{1O}$, $TC_{2O}$, $CB_{1O}$, $CB_{2F}$]\\
	\hspace{2mm} $Path_2$  = [$CT_O$, $R_O$, $TC_{1O}$, $TC_{2O}$, $CB_{1F}$, $CB_{2O}$]\\
	\hspace{2mm} $Path_3$  = [$CT_O$, $R_O$, $TC_{1O}$, $TC_{2O}$, $CB_{1F}$, $CB_{2F}$]\\
	\hspace{2mm} $Path_4$  = [$CT_O$, $R_O$, $TC_{1O}$, $TC_{2F}$, $CB_{1O}$]\\
	\hspace{2mm} $Path_5$  = [$CT_O$, $R_O$, $TC_{1O}$, $TC_{2F}$, $CB_{1F}$]\\
	\hspace{2mm} $Path_6$  = [$CT_O$, $R_O$, $TC_{1F}$, $TC_{2O}$, $CB_{2O}$]\\
	\hspace{2mm} $Path_7$  = [$CT_O$, $R_O$, $TC_{1F}$, $TC_{2O}$, $CB_{2F}$]\\
	\hspace{2mm} $Path_8$  = [$CT_O$, $R_O$, $TC_{1F}$, $TC_{2F}$]\\
	\hspace{2mm} $Path_9$  = [$CT_O$, $R_F$]\\
	\hspace{2mm} $Path_{10}$ = [$CT_F$]\\
\end{flushleft}

\noindent\textit{Step 3 (Partition Outcome Space)}: \\
The partitioning of the outcome space is essential as we are only interested in the occurrence of certain events in an ET. Suppose, we are only focusing on the failure of $CB_1$, then paths 2, 3, and~5-10 are obtained. Similarly, different sets of paths can be obtained by observing the behavior of the trip circuit components as:
\begin{itemize}
	\item $\mathcal{P}$ ($CB_{1}$ Only Fails) = $\sum \mathcal{P} (2, 3, 5-10)$
	\item $\mathcal{P}$ ($CB_{1}$ Only Operates) = $\sum \mathcal{P} (0, 1, 4)$
	\item $\mathcal{P}$ ($CB_{2}$ Only Fails) = $\sum \mathcal{P} (1, 3-5, 7-10)$
	\item $\mathcal{P}$ ($CB_{2}$ Only Operates) = $\sum \mathcal{P} (0, 2, 6)$
	\item $\mathcal{P}$ (Both $CB_{1}$ and $CB_{2}$ Fail) =  $\sum \mathcal{P} (3, 5, 7-10)$
	\item $\mathcal{P}$ (Both $CB_{1}$ and $CB_{2}$ Operate) = $\sum \mathcal{P} (0)$
\end{itemize} 
\noindent To the best of our knowledge this feature is not found in any other existing ET analysis tool.\\

\noindent\textit{Step 4 (Probability Evaluation)}: \\
To estimate the probability of events associated with the trip circuit components, we assign probability values to each operational state of the components, as shown in Table~\ref{Table: Trip circuit components States}. Assume that the times to failure of the trip circuit components are exponentially distribution with failure rate $\lambda$ and time index \textit{t}. Then the unreliability function or the probability of failure can be computed as \cite{allan2013reliability}: 
\begin{equation}
\centering 
F(t) = \mathcal{P} (X \leq t) = 1 - e^{-\lambda t}
\end{equation}
where \textit{X} is a time-to-failure random variable. Similarly, the reliability of a component can be estimated by taking the complement of unreliability function with respect to the probability space as \cite{allan2013reliability}: 
\begin{equation}
\centering 
R(t) = \mathcal{P} (X > t) = 1 - F(t) 
\end{equation}
The probabilities of the different trip circuit events, which are calculated using $\mathcal{ETMA}$ are as follows: \\
\noindent $\mathcal{P}$ (Both $CB_{1}$ and $CB_{2}$ Fail) = 5.389960806400000\% \\ 
\hspace{-2mm}  $\mathcal{P}$ (Both $CB_{1}$ and  $CB_{2}$ Operate) = 82.429704806399980\%  \\
\hspace{-2mm}  $\mathcal{P}$ ($CB_{1}$ Only Fails) = 11.480127999999999\% \\
\hspace{-2mm}  $\mathcal{P}$ ($CB_{1}$ Only Operates) = 88.519871999999980\%  \\
\hspace{-2mm}  $\mathcal{P}$ ($CB_{2}$ Only Fails) = 11.480127999999999\%  \\
\hspace{-2mm}  $\mathcal{P}$ ($CB_{2}$ Only Operates) = 88.519871999999980\% \\

\begin{table}[!t]
	\caption{\protect  \centering Trip circuit probability of components states}
	\label{Table: Trip circuit components States}
	\begin{center}
		\begin{tabular}{|c|c|c|c|} 
			\hline \thead{Component} & \thead{ \makecell{ $\lambda$ \\ \textit{(f/yr)}}}   & \thead{Prob. of Failure (\%) \\ After 6 Months} & \thead{Prob. of Success (\%) \\ After 6 Months} \\ [0.5ex] 
			\hline\hline
			\makecell{CT} & 0.06  & \makecell{$CT_F$} (3\%) & \makecell{$CT_O$} (97\%)\\  [0.5ex]
			\hline
			\makecell{R} & 0.04  &\makecell{$R_F$} (2\%) & \makecell{$R_O$} (98\%)\\ [0.5ex]
			\hline
			\makecell{$TC_1$} & 0.08   & \makecell{$TC_{1F}$} (4\%) & \makecell{$TC_{1O}$} (96\%) \\ [0.5ex]
			\hline
			\makecell{$TC_2$} & 0.08  & \makecell{$TC_{2F}$} (4\%) & \makecell{$TC_{2O}$} (96\%)\\ [0.5ex]
			\hline
			\makecell{$CB_1$} & 0.06  & \makecell{$CB_{1F}$} (3\%) & \makecell{$CB_{1O}$} (97\%) \\ [0.5ex]
			\hline
			\makecell{$CB_2$}  & 0.06  & \makecell{$CB_{2F}$} (3\%) & \makecell{$CB_{2O}$} (97\%)\\ [0.5ex]
			\hline
		\end{tabular}
	\end{center}
\end{table}

It can be observed that the probability of both circuit breakers $CB_1$ and $CB_2$ failing is evaluated as 5.389960806400000\%. If we want to decrease their probability to 2.5\% or less, then we may add redundancy to these components. However, to ensure that the redundancy to these components are a correct decision, we need to conduct the decision analysis of the trip circuit, which is presented in the next section.


\subsection{Decision Analysis}

In the trip circuit, we can identify that the critical components are  CT and R since the failure of these components may cause overall trip circuit failure. A decision-tree describing the process of selecting the redundancy for critical trip circuit components is shown in Fig. \ref{Fig: Decision tree for CT duplication}. First, we select CT only for redundancy (i.e., adding $CT_2$) assuming the same probability of failure and success of $CT_1$. If the probability of both circuit breakers $CB_1$ and $CB_2$ failing together, after re-evaluation in $\mathcal{ETMA}$, is equal to 2.5 \% or less as required, then this is a correct decision. If not, then we select the critical component R for redundancy. If we still do not achieve the desired level of probability, then we select both CT and R together. If the results are not acceptable, then we make a new component selection from the trip circuit. We use $\mathcal{ETMA}$ to generate the new reduced ET model after adding redundant $CT_2$  and obtain the following event outcome space (31 paths only out of 128 complete paths):

\begin{figure}[!t]
	\begin{center}
		\centering
		\includegraphics[width = 0.8 \columnwidth]{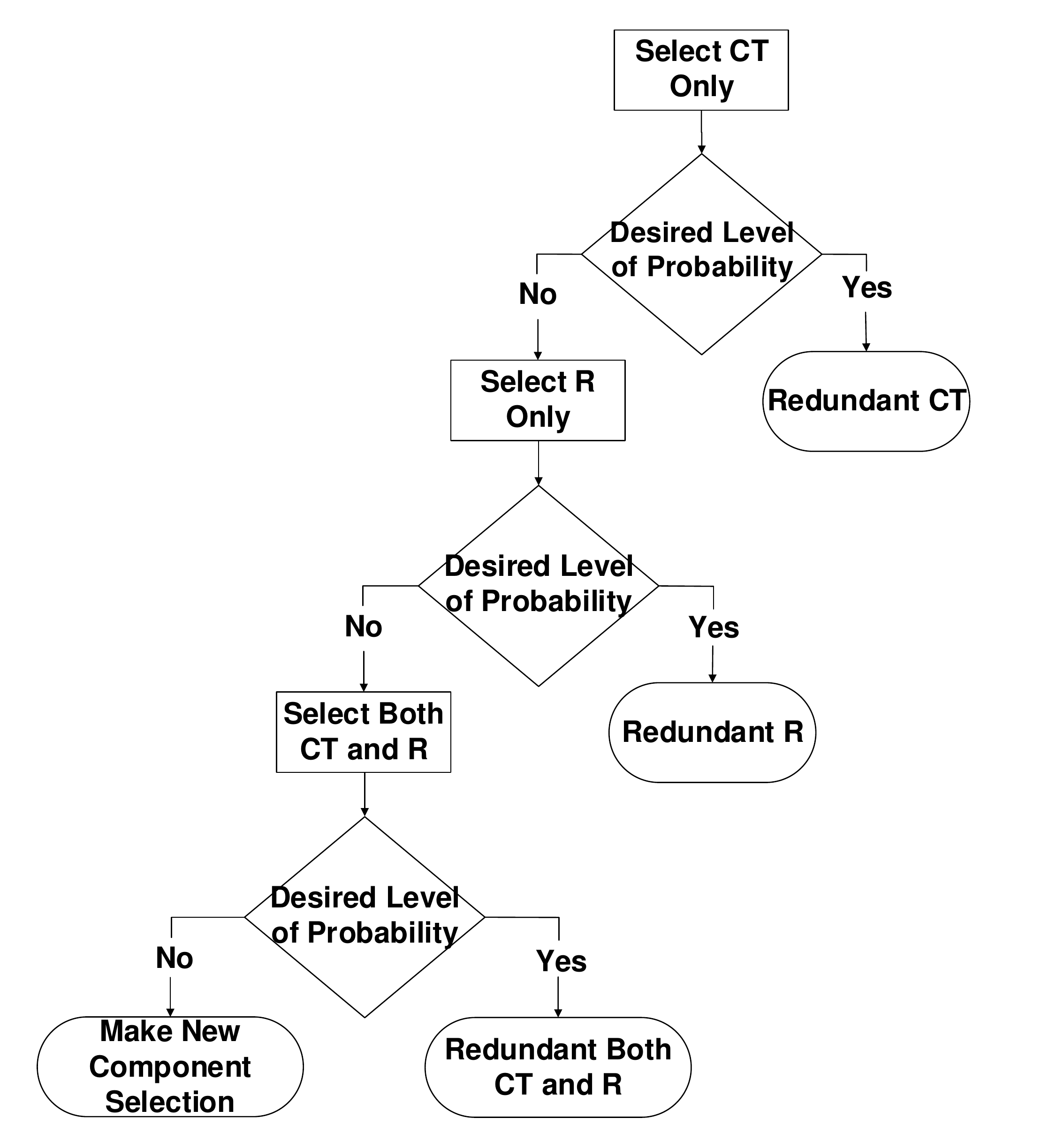} 
	\end{center}
	\caption{\protect Decision tree for the trip circuit}
	\label{Fig: Decision tree for CT duplication}
\end{figure}

\begin{flushleft}
	$Path_0$ = [$CT_{1O}$, $CT_{2O}$, $R_O$, $TC_{1O}$, $TC_{2O}$, $CB_{1O}$, $CB_{2O}$] \\
	$Path_1$ = [$CT_{1O}$, $CT_{2O}$, $R_O$, $TC_{1O}$, $TC_{2O}$, $CB_{1O}$, $CB_{2F}$]\\
	$Path_2$ = [$CT_{1O}$, $CT_{2O}$, $R_O$, $TC_{1O}$, $TC_{2O}$, $CB_{1F}$, $CB_{2O}$]\\
	$Path_3$ = [$CT_{1O}$, $CT_{2O}$, $R_O$, $TC_{1O}$, $TC_{2O}$, $CB_{1F}$, $CB_{2F}$]\\
	$Path_4$ = [$CT_{1O}$, $CT_{2O}$, $R_O$, $TC_{1O}$, $TC_{2F}$, $CB_{1O}$]\\
	$Path_5$ = [$CT_{1O}$, $CT_{2O}$, $R_O$, $TC_{1O}$, $TC_{2F}$, $CB_{1F}$]\\
	$Path_6$ = [$CT_{1O}$, $CT_{2O}$, $R_O$, $TC_{1F}$, $TC_{2O}$, $CB_{2O}$]\\
	$Path_7$ = [$CT_{1O}$, $CT_{2O}$, $R_O$, $TC_{1F}$, $TC_{2O}$, $CB_{2F}$]\\
	$Path_8$ = [$CT_{1O}$, $CT_{2O}$, $R_O$, $TC_{1F}$, $TC_{2F}$]\\
	$Path_9$ = [$CT_{1O}$, $CT_{2O}$, $R_F$]\\
	$Path_{10}$ = [$CT_{1O}$, $CT_{2F}$, $R_O$, $TC_{1O}$, $TC_{2O}$, $CB_{1O}$, $CB_{2O}$]\\
	$Path_{11}$ = [$CT_{1O}$, $CT_{2F}$, $R_O$, $TC_{1O}$, $TC_{2O}$, $CB_{1O}$, $CB_{2F}$]\\
	$Path_{12}$ = [$CT_{1O}$, $CT_{2F}$, $R_O$, $TC_{1O}$, $TC_{2O}$, $CB_{1F}$, $CB_{2O}$]\\
	$Path_{13}$ = [$CT_{1O}$, $CT_{2F}$, $R_O$, $TC_{1O}$, $TC_{2O}$, $CB_{1F}$, $CB_{2F}$]\\
	$Path_{14}$ = [$CT_{1O}$, $CT_{2F}$, $R_O$, $TC_{1O}$, $TC_{2F}$, $CB_{1O}$]\\
	$Path_{15}$ = [$CT_{1O}$, $CT_{2F}$, $R_O$, $TC_{1O}$, $TC_{2F}$, $CB_{1F}$]\\
	$Path_{16}$ = [$CT_{1O}$, $CT_{2F}$, $R_O$, $TC_{1F}$, $TC_{2O}$, $CB_{2O}$]\\
	$Path_{17}$ = [$CT_{1O}$, $CT_{2F}$, $R_O$, $TC_{1F}$, $TC_{2O}$, $CB_{2F}$]\\
	$Path_{18}$ = [$CT_{1O}$, $CT_{2F}$, $R_O$, $TC_{1F}$, $TC_{2F}$]\\
	$Path_{19}$ = [$CT_{1O}$, $CT_{2F}$, $R_F$]\\
	$Path_{20}$ = [$CT_{1F}$, $CT_{2O}$, $R_O$, $TC_{1O}$, $TC_{2O}$, $CB_{1O}$, $CB_{2O}$]\\
	$Path_{21}$ = [$CT_{1F}$, $CT_{2O}$, $R_O$, $TC_{1O}$, $TC_{2O}$, $CB_{1O}$, $CB_{2F}$]\\
	$Path_{22}$ = [$CT_{1F}$, $CT_{2O}$, $R_O$, $TC_{1O}$, $TC_{2O}$, $CB_{1F}$, $CB_{2O}$]\\
	$Path_{23}$ = [$CT_{1F}$, $CT_{2O}$, $R_O$, $TC_{1O}$, $TC_{2O}$, $CB_{1F}$, $CB_{2F}$]\\
	$Path_{24}$ = [$CT_{1F}$, $CT_{2O}$, $R_O$, $TC_{1O}$, $TC_{2F}$, $CB_{1O}$]\\
	$Path_{25}$ = [$CT_{1F}$, $CT_{2O}$, $R_O$, $TC_{1O}$, $TC_{2F}$, $CB_{1F}$]\\
	$Path_{26}$ = [$CT_{1F}$, $CT_{2O}$, $R_O$, $TC_{1F}$, $TC_{2O}$, $CB_{2O}$]\\
	$Path_{27}$ = [$CT_{1F}$, $CT_{2O}$, $R_O$, $TC_{1F}$, $TC_{2O}$, $CB_{2F}$]\\
	$Path_{28}$ = [$CT_{1F}$, $CT_{2O}$, $R_O$, $TC_{1F}$, $TC_{2F}$]\\
	$Path_{29}$ = [$CT_{1F}$, $CT_{2O}$, $R_F$]\\
	$Path_{30}$ = [$CT_{1F}$, $CT_{2F}$]\\
\end{flushleft}

\noindent The new probabilities values evaluated using $\mathcal{ETMA}$ describing the occurrence of failure and success in the trip circuit components are as~follows:

\noindent $\mathcal{P}$ (Both $CB_{1}$ and $CB_{2}$ Fail) = 2.255165963059199\%\\
\hspace{-2mm} $\mathcal{P}$ (Both $CB_{1}$ and $CB_{2}$ Operate) = 84.902595950591990\%\\
\hspace{-2mm} $\mathcal{P}$ ($CB_1$ Only Fails) = 8.824531840000000\%\\
\hspace{-2mm} $\mathcal{P}$ ($CB_1$ Only Operates) = 91.175468160000000\%\\
\hspace{-2mm} $\mathcal{P}$ ($CB_2$ Only Fails) = 8.824531840000000\%\\
\hspace{-2mm} $\mathcal{P}$ ($CB_2$ Only Operates) = 91.175468160000000\%\\

\begin{figure}[!h]
	\begin{center}
		\centering
		\includegraphics[width = 0.78 \columnwidth]{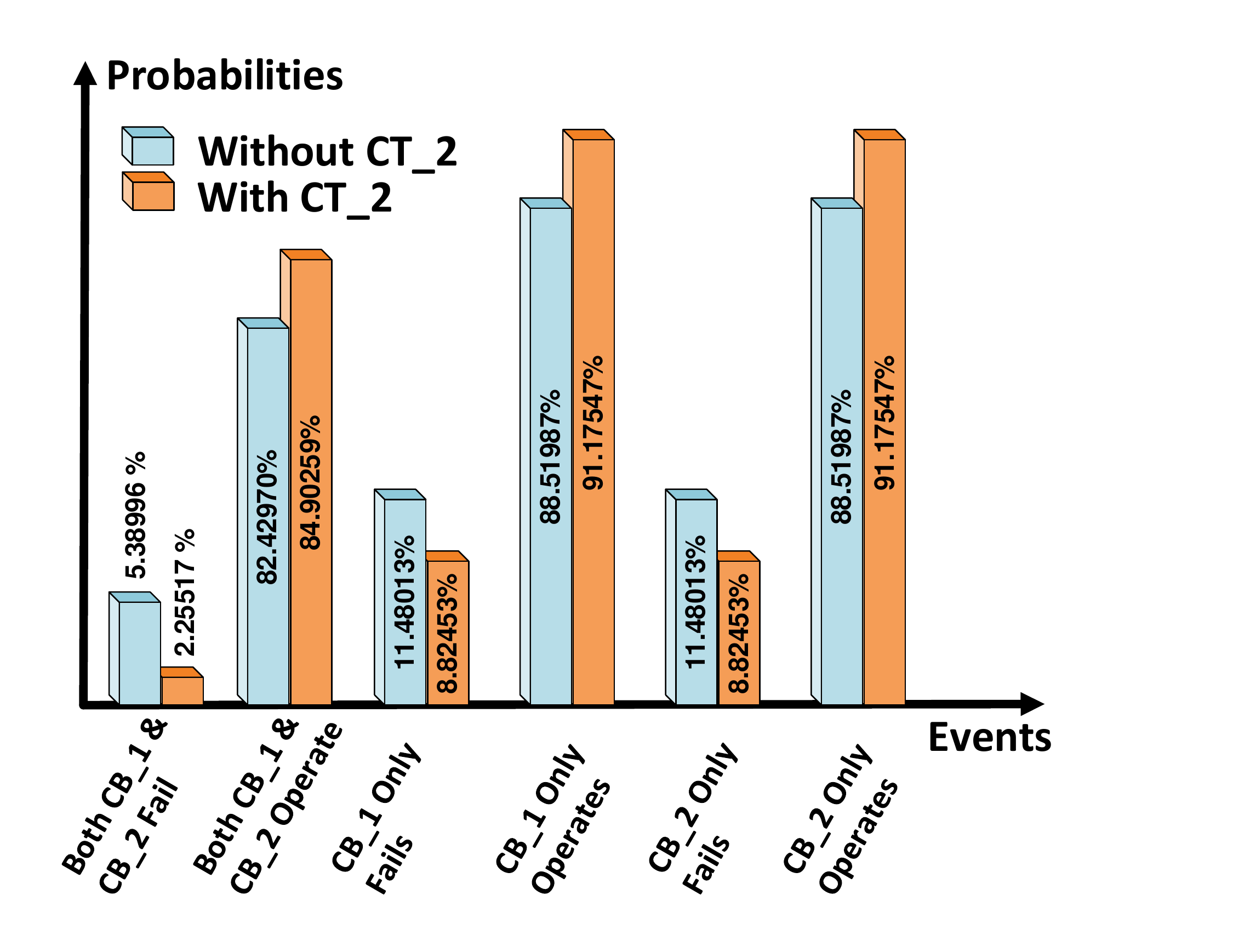} 
	\end{center}
	\caption{\protect Trip circuit events probabilities evaluation}
	\label{Fig: Trip circuit reliability analysis}
\end{figure}

By comparing these values with those in Section~\ref{case study} (\textit{Step}~4), we can clearly observe that the trip circuit performance has been improved. Fig.~\ref{Fig: Trip circuit reliability analysis} shows the comparison among these values in a histogram plot. It can be seen that the  probability percentage of the circuit breakers $CB_1$ and $CB_2$ failing together is decreased from 5.38996\% to 2.25517\% by an amount of 3.13479\%. Similarly, the proportion of the circuit breakers $CB_1$ and $CB_2$ succeeding together is also increased from 82.42970\% to 84.90259\% with an increment of 2.47289\%. 

\section{Comparison with Isograph}

\begin{figure}[!h]
	\begin{center}
		\centering
		\includegraphics[width = 0.85 \columnwidth]{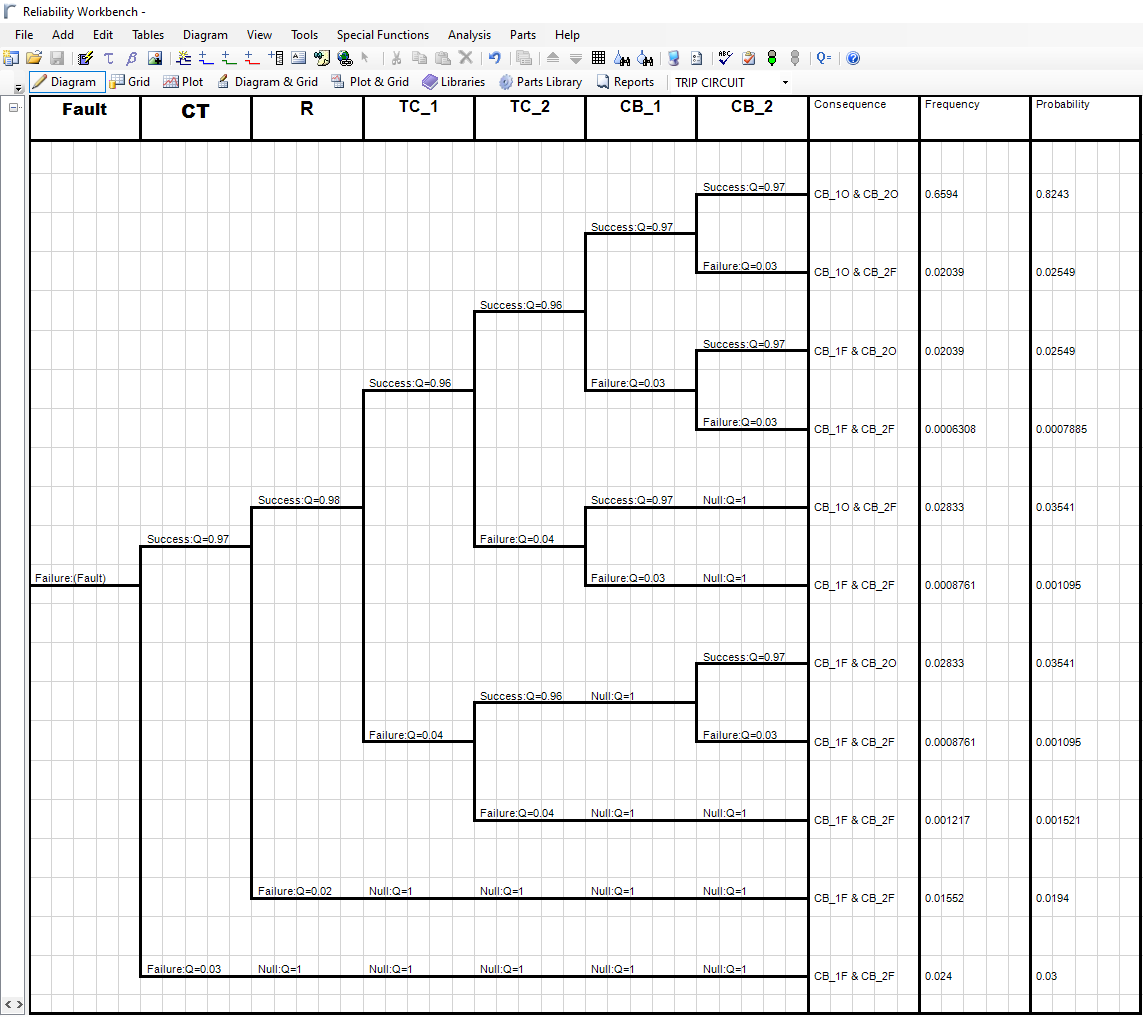}
	\end{center}
	\caption{\protect \centering Isograph: Trip circuit ET model}
	\label{Fig: manually drawn of Trip circuit ET model}
\end{figure}

To ensure the accuracy of the $\mathcal{ETMA}$ computation, we compare the trip circuit analysis results with the commercial Isograph ET analysis tool \cite{Isograph_tp}. We analyze the trip circuit without any redundancy in the critical components using Isograph. It is important to mention that, unlike  $\mathcal{ETMA}$, Isograph requires from the users to manually draw the trip circuit actual ET model ($\mathcal{ETMA}$ \textit{Step} 2) and assign the probability to each event as shown in Fig. \ref{Fig: manually drawn of Trip circuit ET model}. Since the partitioning process of the ET paths is not available in Isograph, we used the manual calculation of the paths probabilities that represent the occurrence of the the trip circuit events. The comparison in the probabilistic analysis of the trip circuit between $\mathcal{ETMA}$ and Isograph is presented in Table \ref{Table: Isograph comparison}. \\

\begin{table} [!h]
	\caption{\protect \centering Comparison between $\mathcal{ETMA}$ and Isograph}
	\label{Table: Isograph comparison}
	\begin{center}
		\begin{tabular}{|l|c|c|} 
			\hline \thead{\makecell{Trip Circuit \\ Events}} & \thead{\makecell{\% Prob. \\ from \\ Isograph}} & \thead{\makecell{\% Prob. \\ from \\  $\mathcal{ETMA}$}} \\ [0.5ex] 
			\hline\hline
			Both $CB_{1}$ and $CB_{2}$ Fail & \makecell{5.38996 \%}  & \makecell{5.389960806400000 \%} \\ [0.5ex]
			\hline
			Both $CB_{1}$ and $CB_{2}$ Operate & \makecell{82.43 \%} & \makecell{82.429704806399980 \%} \\ [0.5ex]
			\hline
			$CB_{1}$ Only Fails & \makecell{11.48 \%}  & \makecell{11.480127999999999 \%} \\ [0.5ex]
			\hline
			$CB_{1}$ Only Operates & \makecell{88.52 \%} & \makecell{88.519871999999980 \%} \\ [0.5ex]
			\hline
			$CB_{2}$ Only Fails & \makecell{11.48 \%} & \makecell{11.480127999999999 \%} \\ [0.5ex]
			\hline
			$CB_{2}$ Only Operates  & \makecell{88.52 \%} & \makecell{88.519871999999980 \%} \\ [0.5ex]
			\hline
		\end{tabular}
	\end{center}
\end{table}

\begin{table}[!h]
	\caption{\protect \centering $\mathcal{ETMA}$: Trip circuit CPU time}
	\label{Table: Trip circuit CPU time}
	\begin{center}
		\begin{tabular}{|c|c|c|c|c|c|} 
			\hline \thead{\makecell{\textit{Steps}}} & \thead{ \makecell{CPU \\ Time \\ $\mathcal{ETMA}$ \\(\textit{Seconds})}}  & \thead{ \makecell{CPU \\ Time \\ Isograph \\(\textit{Seconds})}} & \thead{\makecell{\textit{Steps}}} & \thead{ \makecell{CPU \\ Time \\ $\mathcal{ETMA}$ \\(\textit{Seconds})}} & \thead{ \makecell{CPU \\ Time \\ Isograph \\(\textit{Seconds})}} \\ [0.5ex] 
			\hline\hline
			\makecell{\textit{Step} 1} & 0.291600  & NA &   \makecell{\textit{Step} 3} & 0.000631 & NA \\  [0.5ex]
			\hline
			\makecell{\textit{Step} 2} & 0.000162 & NA &   \makecell{\textit{Step} 4} & 0.004319 & 2.752 \\  [0.5ex]
			\hline
		\end{tabular}
	\end{center}
\end{table}

It can be observed that the probabilities obtained from  $\mathcal{ETMA}$ are approximately equivalent to the corresponding ones calculated using Isograph. This clearly demonstrates that $\mathcal{ETMA}$ is not only providing the correct results but also a complete ET analysis compared to existing ET analysis tools. Moreover, the CPU time for the trip circuit step-wise ET analysis in $\mathcal{ETMA}$ is much faster than Isograph, as shown in Table \ref{Table: Trip circuit CPU time}.  The experiments were performed on a single-core i5, 2.20 GHz, Linux VM with 1 GB of RAM device. Also, $\mathcal{ETMA}$ is providing several additional features, including the automation of complete ET generation and the partitioning of ET paths for events probabilistic analysis, that are not available in any other existing reliability analysis tool.  All these features of  $\mathcal{ETMA}$ are extremely useful for safety analysts and reliability engineers to quantify system improvements with fast and accurate decisions. 

\section{Conclusions}

In this paper, we proposed a new event trees modeling and analysis tool, called $\mathcal{ETMA}$, using \textit{list} data-structure in Python. $\mathcal{ETMA}$ provides several features to model any generic, complete and sequential ET diagram consisting of a large number of system components. Also, $\mathcal{ETMA}$ provides deleting/reducing features to remove irrelevant specific nodes or paths from a complete ET diagram to model the exact logical behavior of the given system. Moreover, $\mathcal{ETMA}$ provides partitioning of ETs paths and probabilistic analysis of the occurrence of a certain event. For illustration purposes, we conducted the ET modeling and analysis of the trip circuit in the power grid transmission lines. The results of $\mathcal{ETMA}$ were used for making redundancy decisions about the critical components in the trip circuit. We also compared the results obtained in $\mathcal{ETMA}$ with those from the Isograph tool, which is commonly used for ET analysis. We plan to extend our $\mathcal{ETMA}$ tool with additional features for safety assessment \cite{qiang2008approach}, reliability analysis \cite{tavner2007reliability} and machine learning~\cite{raschka2017python}.


\bibliographystyle{IEEEtran}	
\bibliography{Technical_Report2}

\end{document}